\documentstyle{l-aa}     
\input psfig   
\newcommand{\asec}  {\hbox{$^{\prime\prime}$}}

\begin{document}
\topmargin=+2.0cm
\thesaurus{ 
	    09.01.1;  
            09.07.1;  
            09.13.2;  
            11.17.1;  
            13.19.3   
	         }
   \title{ New upper limits on the interstellar O$_2$ abundance}
   \subtitle{}
   \author{F.~Combes\inst{1}, T.~Wiklind\inst{2}, N.~Nakai\inst{3}}
 \offprints{F.~Combes, bottaro@obspm.fr}        
   \institute{ Observatoire de Paris, DEMIRM, 61 av. de l'Observatoire, 
F-75014 Paris
\and  Onsala Space Observatory, S--43992 Onsala, Sweden  
\and  Nobeyama Radio Observatory, Minamimaki-mura, Minamisaku-gun, Nagano
384-13, Japan}
   \date{Received date; Accepted date}
   \maketitle
%
   \begin{abstract}
We report new observations of molecular oxygen in absorption at $z=0.685$
in front of the radio source B0218+357. The lines at 56.3 and 118.7 GHz
have been observed, redshifted to 33.4 and 70.5 GHz respectively, with the
12m at Kitt Peak, 43m at Green Bank telescopes, and the 45m Nobeyama
radio telescope. Deriving the surface filling factor of the absorbing
dark cloud with other lines detected at nearby frequencies, we deduce from
the upper limits on the O$_2$ lines a relative abundance of molecular
oxygen with respect to carbon monoxyde of O$_2$/CO $\la$ 2 10$^{-3}$ at
1$\sigma$, seven times lower than the previous limit.  The consequences of
this result are discussed.
   \end{abstract}
\section{Introduction}
The interstellar abundance of molecular oxygen has been a puzzle
for a long time. Since the cosmic abundance of oxygen (O/H $\approx$ 8.5
10$^{-4}$) is a little more than twice as large as that of carbon, at most 
half of oxygen could be locked up in CO, the rest should be in the form of
atomic OI, or in the molecules O$_2$, OH, H$_2$O.
 The H$_2$O/H$_2$ ratio is predicted between 10$^{-7}$ and 10$^{-5}$ 
by chemical models, and such predictions are compatible with
recent ISO results (e.g. Cernicharo et al 1997). Since the abundance
of OH is only 10$^{-6}$ or lower, it was long thought that O$_2$
could be as abundant as CO, i.e. 10$^{-5}$-10$^{-4}$ with respect
to H$_2$ or at least 1/10th of that of CO (Langer et al 1984, Viala 1986).
 In dark clouds, OI and O$_2$ are predicted to have comparable abundances,
while most of the oxygen should be atomic in diffuse clouds.
 To circumvent the atmospheric absorption, the recent searches have
used either the $^{16}$O$^{18}$O (Fuente et al 1993, Mar\'echal et al 1997a)
or the redshifted O$_2$ (Combes et al 1991, Liszt 1992, Combes \& Wiklind
1995) and have come up with an upper limit in the O$_2$/CO ratio
of $\approx$ 1.3  10$^{-2}$ at 1$\sigma$. 

We report about further efforts to detect an O$_2$ line in absorption,
using the redshifted molecular cloud at $z=0.68466$ absorbing the
continuum from the quasar B0218+357 (cf Combes \& Wiklind 1995). The selected
lines are at lower frequencies, where the continuum flux is larger,
and should provide a better chance of detection.

\section {Observations}

The observations around 70 GHz have been carried out at the NRAO
12--m telescope at Kitt Peak\footnote{NRAO is a facility of
the National Science Foundation, operated under cooperative agreement
by Associated Universities Inc.} in February 1996.
The observations around 30 GHz have been obtained with the NRAO
43m telescope in Green Bank in February 1996, and then at the 45m
Nobeyama telescope in Japan during several observing runs
from February to May 1997. 

\smallskip
{\it \bf Kitt-Peak 12--m}
Two SIS receivers (68-90 and 90-116 GHz) were used at 70 GHz (O$_2$), 
87 GHz (CS(3--2)) and 105 GHz (HCN(2--1)), with an average
system temperature of 345, 225 and 250 K respectively. 
 The receivers were used in dual polarization modes, and tuned in single-
sideband.
The corresponding beam sizes are 98, 79, and 66 \asec. 
The observations were done using the wobbler switching mode.
We used simultaneously broad filterbanks (0.5 MHz resolution) and narrow band
autocorrelators (hybrid spectrometer), providing a resolution of 97 KHz.
 However the spectra were smoothed to gain signal-to-noise to velocity
resolutions of 1-3 km/s.
Pointing was checked at least every two hours by broadband continuum
observations of radio-sources;
the pointing accuracy was 6 \asec\,  rms.

\smallskip
{\it \bf Green-Bank 43--m}
The receiver was a cooled HFET with  a system temperature T$_{sys}$ = 120K,
and a bandwidth of 500 MHz. The beam size is 56 \asec\, at the CS and O$_2$
frequencies. The conversion ratio was estimated at 10 Jy/K. Unfortunately,
the efficiency was strongly varying with the hour angle, up to a factor
2-3. The efficiency was monitored through observations of strong continuum
sources (pointing sources) such as 3C 274, 3C 84, and the pointing corrections
were also obtained on Saturn and Jupiter. 
 The observations were made by position switching. 
The autocorrelator backend provided a resolution of
19.5 KHz, or 0.18 km/s at the O$_2$ frequency. 

\smallskip
{\it \bf Nobeyama 45--m}
The receiver was a cooled HEMT of bandwidth 2GHz and T$_{sys}$ 140--170K.
The half-power beam width of the telescope is 59 \asec\, and 51 \asec\, at  
29 and 33 GHz respectively. The aperture efficiency is 60\% at 30 GHz,
leading to a conversion ratio of 2.9 Jy/K, for the antenna temperature T$_A^*$.
  The observing procedure was position switching. The 
was checked every one to two hours by observing a nearby SiO maser star,
S per, and its accuracy was better than 10 \asec\, (peak value).
 The 2048-chanels AOS backends provide a resolution of 37 kHz,
i.e. 0.38km/s at the CS frequency and 0.33 GHz at the O$_2$ one.
The flux density of the continuum emission was measured with a continuum
backend in beam-switching mode at 29 GHz and with AOS at 33 GHz
simultaneously with the O$_2$ line.

\begin{figure}
\psfig{figure=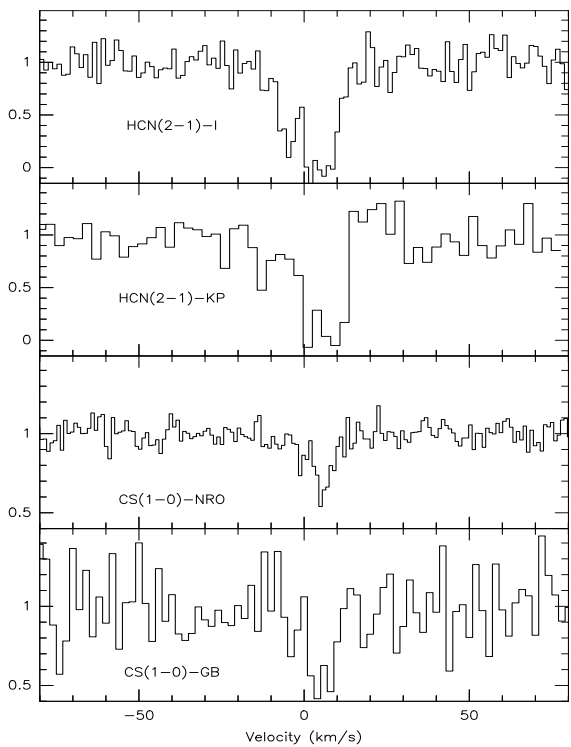,bbllx=30mm,bblly=25mm,bburx=200mm,bbury=270mm,height=10.cm,width=8.cm}  
\caption[]{ From top to bottom:
HCN(2--1) spectrum at IRAM July 1996, HCN(2--1) at Kitt-Peak February 1996;
CS(1--0) at Nobeyama February-May 1997, CS(1--0) at Green-Bank February 1996.
Spectra have been normalised to the continuum seen by
the molecular cloud, i.e. 33\% of the total continuum.}
\label{figo1}
\end{figure}

\section {Results } 
The BL Lac object B0218+357 is remarkable by its radio 
images, that reveal two distinct flat-spectrum cores (A and B component),
with similar spectrum and polarization level, strongly suggesting
a gravitational lens phenomenon (Patnaik et al 1995). That A and B are 
gravitational images is confirmed by the presence of a small
 Einstein ring surrounding the B image, of 335 milli-arcsecond in 
diameter (Patnaik et al 1993). Since the ring has a steeper spectrum, it is 
probably the image of a hot spot in the jet component, and its contribution
in the millimetric domain should be negligible.
 The lensing galaxy at a redshift $z=0.68466$ absorbs the radio continuum of 
the remote quasar in the HI line (Carilli et al 1993), and in molecular lines
(Wiklind \& Combes 1995).
The intensity ratio between the two images is A/B $\approx$ 3-4 at several 
radio wavelengths (O'Dea et al. 1992, Patnaik et al. 1993).
  As in the similar absorbing lens galaxy in front of PKS1830--211
(cf Wiklind \& Combes 1996), some of the main molecular lines are optically
thick as attested by the detection of isotopomers, but the radio continuum
is not totally absorbed, which means that 
the absorbing material does not cover the whole
surface of the continuum source. In the case of PKS1830--211, interferometric
observations have confirmed that only one image of
the source is covered by the absorbing cloud (Wiklind \& Combes 1997a).
 In B0218+357, the situation is still unclear, although it has been
claimed that only part of the A image is covered
at cm wavelengths (Menten \& Reid 1996) and through optical imaging
(Grundahl \& Hjorth 1995). At the period of our observations,
the fraction of the total continuum absorbed was determined to be 
one third. 

\begin{figure}
\psfig{figure=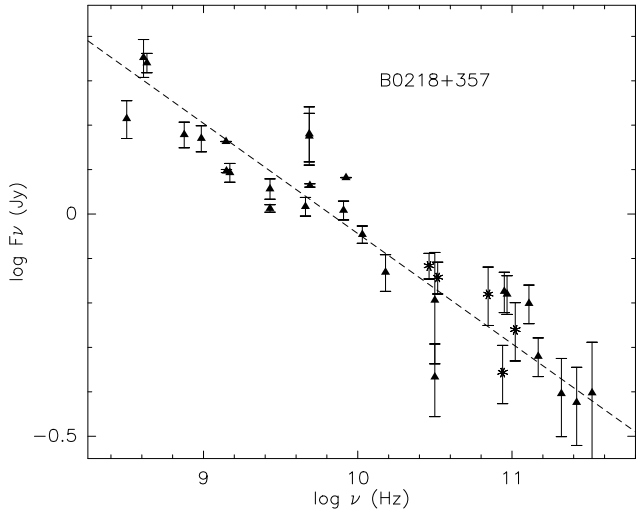,bbllx=10mm,bblly=5mm,bburx=205mm,bbury=235mm,height=7.cm,width=8.5cm,angle=-90} 
\caption[]{ Radio continuum spectrum of B0218+357. Measurements from
the literature (see NED for a compilation) are  plotted with filled triangles, 
and the stars correspond to the present work.
The dashed line is not a fit, but represents a power law of slope $-0.25$. }
\label{figo2}
\end{figure}

To derive the optical depth of a given line, or at least an upper limit on the 
optical depth, we must know exactly the total continuum flux of the source 
at the epoch of observations, since it might be variable, and the surface 
filling factor of the absorption, i.e. the fraction of the continuum source 
surface covered by molecular clouds. The best is even to bypass these 
two steps, by observing another line at a neighbouring frequency, which is
optically thick and gives directly the absorbed continuum level. This was 
done in our previous work, where the neighbouring CO transition was 
shown to be highly optically thick through observations of $^{13}$CO and
C$^{18}$O (Combes \& Wiklind 1995).

\begin{figure}
\psfig{figure=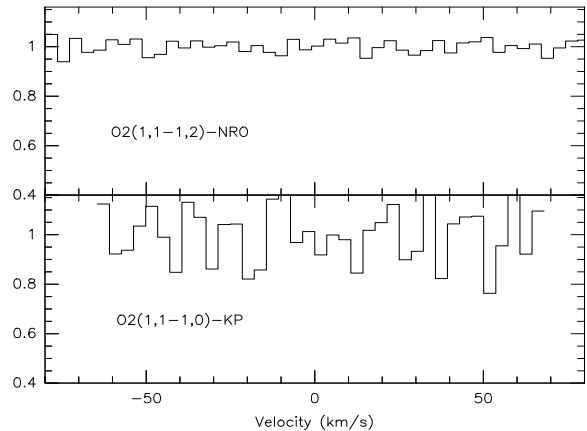,bbllx=23mm,bblly=135mm,bburx=200mm,bbury=270mm,height=6.cm,width=8.5cm} 
\caption[]{ Negative results found for the two selected O$_2$ lines:
{\it Top} : the N,J = 1,1 -- 1,2 line observed at Nobeyama at 33 GHz;
{\it Bottom} : the N,J = 1,1 -- 1,0 line observed at Kitt Peak at 70 GHz.
Spectra have been normalised to the continuum seen by
the molecular cloud, i.e. 33\% of the total continuum. The velocity resolution
is smoothed for both spectra to 3.6 km/s.}
\label{figo3}
\end{figure}

To calibrate   the N,J = 1,1 -- 1,0 118 GHz O$_2$ line, redshifted at 70 
GHz, we could not observe the 115 GHz CO(1--0) line unfortunately, since 
the 12--m receiver was not operating at 68 Ghz. Instead, the line CS(3--2) 
and HCN(2--1) were observed at redshifted frequencies of 87 and 105 
GHz respectively. The HCN(2--1) profile is displayed in figure \ref{figo1}, 
normalized to the absorbed
continuum. It can be compared with the analogous spectrum obtained with the 
IRAM telescope in July 1996, also normalised to the continuum. 
Both appear optically thick. From the millimetric observations carried out 
with the IRAM-30--m telescope in 1994, 1995 and 1996, the total 
continuum does not seem to have varied, at least within the error bars 
(10-20\%). The CS(3--2) was not detected with the 12--m (although it
is detected with the IRAM-30--m, Wiklind \& Combes 1997b).

To calibrate the N,J = 1,1 -- 1,2 56 GHz O$_2$ line, redshifted at 33 GHz, 
we have observed the CS(1--0) line, redshifted at 29 Ghz, together with 
the corresponding $^{13}$CS line.  The CS(1--0) line was detected both at NRO 
and GB (see figure \ref{figo1}), but are not saturated, and the $^{13}$CS line 
was not detected, with an rms noise level of 4mK in channels of 4km/s
(see Table 1). We must therefore rely on the
measurement of the total continuum, and on the assumption that the 
fraction of the surface covered by the molecular cloud is the same  
at 1cm and 3mm.

The total continuum level measured at the various frequencies and 
telescopes is displayed in figure \ref{figo2}, together with some other 
measurements in the literature (see Combes \& Wiklind 1997 for details). 
Within the error bars, all measurements follow a power-law distribution as 
a function of frequency. In February 1996, the total continuum level measured 
at 30 GHz at Green-Bank was T$_A^*$ = 70 $\pm$ 10mK (0.7 Jy), and in February 
1997 it was 0.255 $\pm$ .017 K at 33 GHz and 0.245 $\pm$ .020 K at 29 GHz at
Nobeyama (i.e. 0.74 and 0.71 Jy) in very good agreement.

We have assumed thermal equilibrium; this is justified since the absorbing 
molecular cloud is dense with a large optical depth, and the O$_2$ lower levels 
are completely thermalised in these conditions (see Mar\'echal et al 1997b).
Then, to estimate column densities, the usual formula can be used
(cf Combes \& Wiklind 1995), assuming gaussian line profiles, with the 
intrinsic line-width of 4.6 km/s.

\begin{table}
\begin{flushleft}
\caption[]{Observed molecules}
\scriptsize
\begin{tabular}{lcrrcc}
\hline
\multicolumn{1}{c}{Mol.}                     &
\multicolumn{1}{c}{Transition}                   &
\multicolumn{1}{c}{$\nu_{\rm rest}^{a)}$}       &
\multicolumn{1}{c}{$\nu_{\rm obs}$}         &
\multicolumn{1}{c}{Det.$^{b)}$}                  &
\multicolumn{1}{c}{Tel.$^{c)}$}                  \\
\multicolumn{2}{c}{ }                            &
\multicolumn{1}{c}{GHz}                         &
\multicolumn{1}{c}{GHz}                          &
\multicolumn{2}{c}{ }                            \\
\hline
CS             & 0$\rightarrow$1 & 48.990978 &  29.080632 & Yes & NRO \\
CS             & 0$\rightarrow$1 & 48.990978 &  29.080632 & Yes & GB \\
CS             & 2$\rightarrow$3 & 146.969033 &  87.239579  & No & KP \\
$^{13}$CS      & 0$\rightarrow$1 & 46.247567 & 27.452167  & No  & NRO \\
HCN               & 1$\rightarrow$2 & 177.261111 & 105.220700 & Yes & KP \\
\hline
O$_2$    & $1_{0}\rightarrow 1_{1}$ & 118.7503  & 70.489179   & No  & KP \\
O$_2$    & $1_{2}\rightarrow 1_{1}$ & 56.26476  & 33.398288   & No  & GB \\
O$_2$    & $1_{2}\rightarrow 1_{1}$ & 56.26476  & 33.398288   & No  & NRO \\
\hline
\end{tabular}
\ \\
a)\ Rest--frequencies from the JPL catalog (Pickett et al) \\
b)\ Detection: Yes/No. \\
c)\ Telescope: KP$=$Kitt Peak 12--m, GB$=$ Green-Bank 43--m, NRO$=$Nobeyama 45--m.
\end{flushleft}
\end{table}
From the observed profiles of CS(1--0) and CS(3--2), a column density of 
N(CS) of 1.5 10$^{14}$ cm$^{-2}$ is found, with an excitation temperature of 
T$_x$ = 6$^{+3}_{-2}$ K, very close to the background temperature
at the considered redshift T$_{bg0}$ (1+z) = 4.59K. With the previously derived
H$_2$ column density of 5 10$^{23}$  cm$^{-2}$ (Combes \& Wiklind 1995),
the CS abundance is estimated to 0.3 10$^{-9}$. With 
this low abundance, no signal could be expected 
for the $^{13}$CS line.

The O$_2$ profiles at 33 and 70GHz are shown in figure \ref{figo3}. At 33GHz, 
the NRO upper limit is more significant than the Green-Bank one, and we 
show only that one. They correspond to an upper limit of optical depth at 
1$\sigma$ of $\tau \la$ 0.025 and 0.12 respectively. According to the 
adopted excitation temperature, upper limits on the O$_2$ abundance 
are derived in Table 2. They are compared with the column densities
derived from previous observations of the CO molecule (IRAM April 1995), 
assuming unsignificant time variation (Combes \& Wiklind 1995).

The present data together with those obtained earlier 
(Combes \& Wiklind 1995) cover a wide range of conditions in
the absorbing gas, specifically low and high excitation temperatures.
It is therefore not possible to have a large abundance of O$_2$ `hidden'
at certain rotational levels. This is an important aspect of measuring
the O$_2$ abundance since the molecular structure inhibits collisional
excitation of the N,J = 1,0 level from the N,J = 0,0 level (the 118.7
GHz line). This does not apply to radiative transitions (Bergman 1995).
However, we can see in Table 2 that the O$_2$/CO abundance ratio does not
depend much on the adopted temperature. We find for any choice of $T_x$
that the O$_2$/CO ratio is $\le$ 2. 10$^{-3}$ at 1$\sigma$.

\section{Discussion}

The new O$_2$ limit derived above on the abundance ratio 
O$_2$/CO $\le$ 2 10$^{-3}$
begins to constrain significantly the chemical 
models. Mar\'echal et al (1997b) have recently reconsidered the chemistry 
and rotational excitation of the O$_2$ molecules in interstellar clouds, 
either dark, transluscent or diffuse. They have used the most recent 
determinations of reaction rates and collisional cross-sections, and 
explored the effect of UV radiation, density and temperature. They consider 
only steady-state equilibrium. Their conclusions show that the most 
important parameters determining the O$_2$ abundance is the UV 
field, since the O$_2$ molecules are easily photodissociated, and the C/O 
abundance ratio in the gas phase. As soon as C/O is larger than 0.7, the
O$_2$/CO ratio falls below 0.1. In the frame of steady-state chemistry
in a dark cloud, our O$_2$ limit implies a gas phase C/O ratio larger than
1 (at the 1$\sigma$ level) or C/O $\ga$ 0.9 at the 3$\sigma$ level.
 This would mean that most of the oxygen is frozen onto grains.
\begin{table}
\begin{flushleft}
\caption[]{Derived upper limits at 1$\sigma$}
\scriptsize
\begin{tabular}{lcccc}
\hline
\multicolumn{1}{c}{}                     &
\multicolumn{1}{c}{N(O$_2$)}                   &
\multicolumn{1}{c}{N(CO)}                   &
\multicolumn{1}{c}{N(O$_2$)}                   &
\multicolumn{1}{c}{O$_2$/CO}                  \\
\multicolumn{1}{c}{ }                            &
\multicolumn{1}{c}{NRO }                            &
\multicolumn{1}{c}{IRAM }                            &
\multicolumn{1}{c}{KP}                          &
\multicolumn{1}{c}{Best }                            \\
\hline
$\nu$(GHz)   & 56.26476 & 136.845411  & 118.7503 &         \\
$\tau$(1$\sigma$)   & $<$.025 & 1500  &  $<$0.12 &         \\
\hline
T$_x$ = 5K   & $<$1.75e16  &1.51e19  & $<$3.47e16 &   $<$1.16e-3    \\
T$_x$ = 10K   & $<$4.57e16  &2.41e19  & $<$1.08e17 &   $<$1.89e-3    \\
T$_x$ = 15K   & $<$8.90e16  &3.91e19  & $<$2.24e17 &   $<$2.20e-3    \\
\hline
\end{tabular}
\end{flushleft}
\end{table}
Another possibility is to relax the hypothesis of steady-state chemistry,
and consider time-dependent models (Prasad \& Huntress 1980, 
Graedel et al 1982, Langer et al 1984): much more atomic
carbon and oxygen exists at times well before steady-state, at so-called
'early-times', i.e. 1-3 10$^5$ yrs. If steady-state has no time to
establish, because of a short dynamical mixing time between the diffuse
photodissociated phase and the dense dark phase, the relative abundance
of O$_2$ can be reduced by one or two orders of magnitude
(Chi\`eze et Pineau des For\^ets 1989). Most of the oxygen would then
be in atomic form, as supported by the recent detection of [OI] 63$\mu$
in absorption towards Sgr B2 (Baluteau et al 1997). 
Also in the model of Le Bourlot 
et al. (1995) molecular clouds reveal chemical bistability, and 
there may exist a high ionization phase of the ISM with a very low oxygen 
abundance, where the O$_2$/CO ratio is lower by two orders of
magnitude than in the low ionization phase.

 The interstellar abundance of molecular oxygen remains
problematic; balloon and satellite experiments are
much needed. If confirmed, the present low upper limit
 for the O$_2$/CO ratio strongly constrains chemistry models.
If a large fraction of the oxygen abundance is frozen onto grains
we expect enhanced O$_2$ and H$_2$O abundances in regions of shocked
molecular gas.

\acknowledgements
 We are very grateful to the Green Bank and Kitt Peak NRAO engineers and 
operators for their particular care in receiver and band-rejection tunings.
We also thank N. Kuno for his continuum flux measurements at NRO.

\end{document}